\pdfoutput=1
\RequirePackage{fix-cm}
\documentclass[twocolumn,epjc3]{svjour3}
\smartqed
\RequirePackage{graphicx}
\usepackage{amsmath,mathtools,amssymb}
\usepackage{bm}
\usepackage{booktabs}
\usepackage[colorlinks,citecolor=blue,urlcolor=blue,linkcolor=blue]{hyperref}
\hypersetup{pdftitle={Parameter-free bounds on single-scale unitarized Pomeron exchange from the pp diffractive dip},pdfauthor={Ji Li},pdfsubject={Eur. Phys. J. C},pdfkeywords={Pomeron, eikonal, diffractive dip, geometric scaling, unitarity, elastic scattering}}
\newcommand{\dd}{\mathrm d}
\newcommand{\Ein}{\operatorname{Ein}}
\newcommand{\q}{\bm q}
\newcommand{\bb}{\bm b}
\newcommand{\GeVt}{\,\mathrm{GeV}^2}
\newcommand{\GeVm}{\,\mathrm{GeV}^{-2}}
\newcommand{\Tbd}{\mathcal T_{\rm bd}}
\journalname{Eur. Phys. J. C}
\begin{document}

\title{Parameter-free bounds on single-scale unitarized Pomeron exchange from the $pp$ diffractive dip}

\author{Ji Li\thanksref{e1,addr1,addr2,addr3}}

\thankstext{e1}{e-mail: liji24@mails.ucas.ac.cn}

\institute{School of Fundamental Physics and Mathematical Sciences, Hangzhou Institute for Advanced Study,
University of Chinese Academy of Sciences, Hangzhou 310024, China\hypertarget{addr1}{}\label{addr1}
\and Institute of Theoretical Physics, Chinese Academy of Sciences, Beijing 100190, China\hypertarget{addr2}{}\label{addr2}
\and University of Chinese Academy of Sciences, Beijing 100049, China\hypertarget{addr3}{}\label{addr3}}

\date{Received: date / Accepted: date}

\maketitle

\begin{abstract}
In a unitarized amplitude whose opacity has a fixed transverse shape, the width of the opacity is only a length scale, so that
every dimensionless combination of elastic observables depends on the central opacity alone, for any energy
dependence of the parameters.  From the elastic fraction $r=\sigma_{\rm el}/\sigma_{\rm tot}$,
$W=\sigma_{\rm tot}|t_{\rm dip}|/(8\pi)$ and $\Tbd=|t_{\rm bump}|/|t_{\rm dip}|$ we derive parameter-free tests of the
transverse structure of unitarized Pomeron exchange.  For the Gaussian Regge--eikonal $W\le3.75$, whereas the data give
$W=5.3$--6.0 at all eight energies from 23.5~GeV to 13~TeV at which the dip has been located, 7.6 to 71 standard deviations
above the bound, with an energy trend opposite to that of the model.  The energy-independent $\Tbd=1.355\pm0.011$, together
with the rise of $r$ from the ISR to the LHC, excludes a single Gaussian scale independently of $W$.  Neither a free shape
nor a different unitarization removes the conflict: L\'evy opacities with $0.8\le\alpha\le6$ give $\Tbd\ge1.47$ at the
13~TeV value of $r$, and a Gaussian $U$-matrix gives $W=4.17$.  A second transverse scale reproduces $\sigma_{\rm tot}$, $r$,
the forward slope and the dip position at 13~TeV, but with the signature phase the position and the height of the bump cannot
be matched together.  The complex Regge slope commonly used to generate the phase violates $|S(b)|\le1$ beyond
$b_*\simeq6$--8~fm.
\end{abstract}

\section{Introduction}
\label{sec:intro}

The eikonal representation of high-energy small-angle scattering goes back to Glauber's treatment of multiple
scattering~\cite{Glauber} and was derived in quantum field theory by L\'evy and Sucher, who resummed the exchange of
any number of mesons and related the eikonal to Regge behaviour~\cite{LevySucher}.  For hadrons, the exchange of
vacuum quantum numbers is described by the Pomeron, a Regge pole with intercept slightly above one; its single
exchange gives a total cross section rising as $s^\Delta$~\cite{DL}, which violates the Froissart--Martin bound
asymptotically~\cite{Froissart,Martin} and must be supplemented by multi-Pomeron cuts~\cite{Gribov}.  The eikonal
resums these cuts by exponentiating the single-Pomeron opacity in impact-parameter space, and the combination of a
Regge Born term with eikonal or multichannel (Good--Walker) unitarization~\cite{GoodWalker} has become the standard
phenomenology of soft hadronic interactions~\cite{Poghosyan,PDGsoft}.

The diffractive dip was established in $pp$ scattering at the ISR, where its position moved to smaller $|t|$ with
energy and the impact-parameter profile was reconstructed from the data~\cite{AmaldiSchubert}; the filling of the
dip by the real part of the amplitude was analysed early on~\cite{PhillipsBarger}.  Geometric scaling, the
hypothesis that the amplitude depends on $t$ only through $\tau=\sigma_{\rm tot}|t|$, was formulated in the
same period~\cite{DiasDeDeus,BurasDiasDeDeus} and connected the dip position to the total cross section.  At the LHC, TOTEM measured
the differential cross section with a dip at 2.76, 7, 8 and 13~TeV~\cite{TOTEM276,TOTEM7dsdt,TOTEM8dip,TOTEMdsdt}
and the integrated cross sections~\cite{TOTEM7,TOTEM8PRL,TOTEMsig}, and ATLAS made independent measurements at 7, 8
and 13~TeV~\cite{ATLAS7,ATLAS8,ATLAS13}, whose normalization differs from TOTEM by about two standard
deviations~\cite{ATLAS13,PetrovATLAS,PetrovATLAS2}.

Several questions raised by these data remain open.  The forward real part measured at 13~TeV~\cite{TOTEMrho} was
interpreted as a signal of a maximal Odderon~\cite{MartynovNicolescu}, and the comparison with $p\bar p$ data at
1.96~TeV as evidence for Odderon exchange~\cite{D0TOTEM}; Donnachie and Landshoff argued that the phase is tied to
the energy dependence and found $\rho\simeq0.14$ without an Odderon in a multi-energy fit~\cite{DL2019,DL2022}, a
refit with correlated uncertainties gave $\rho=0.11\pm0.01$~\cite{PetrovTkachenko}, a screened Odderon changes $\rho$
by at most 0.004~\cite{LunaRyskinKhoze,LunaRyskinKhoze2}, and the combined significance of the D0--TOTEM result has been
questioned~\cite{CsorgoUniverse}; the status is reviewed in Ref.~\cite{RyskinOdd}.  In impact-parameter space the
13~TeV profile is black at the centre~\cite{Alkin}, possibly beyond the black-disk value, which points to
reflective scattering~\cite{TroshinTyurin,TroshinTyurin2}, and a central depression of the inelastic profile has been reported
\cite{Hollow,CsorgoLevy}.  Unitarization itself is not unique: eikonal and $U$-matrix schemes applied to the same
Born input differ in the strong-absorption regime~\cite{Luna2026,Vanthieghe}.

Realistic models, multichannel eikonals with non-Gaussian form factors, secondary Reggeons and possibly an Odderon,
describe the data over a wide energy range~\cite{KMR,KMR2,KMRslope,HEGS,HEGS2,BroiloSoft}.  Because they are adjusted
through several functions of $s$ and $t$, they do not show why the simplest member of the family fails.  That a
one-channel eikonal cannot reproduce the optical point and the dip position together has been noted for a mini-jet
model~\cite{PancheriGS,PancheriGS2}.

On the empirical side, two scale-free regularities are now established.  The product $\sigma_{\rm tot}|t_{\rm dip}|$
is approximately constant from the ISR to the LHC but differs from its black-disk value
\cite{Bautista,CsorgoNemes,CsorgoNemes2}, and modified scaling laws have been proposed~\cite{PancheriGS,PancheriGS2,KMY,KMY2}.  The ratio of the
bump to the dip position, $\Tbd=1.355\pm0.011$, is independent of energy from 23~GeV to 13~TeV, whereas geometric
scaling of the cross sections holds at the ISR but is violated at the LHC outside the dip--bump
region~\cite{BaldenegroPRS,Praszalowicz,Praszalowicz2,Praszalowicz3,Praszalowicz4}.

Section~\ref{sec:background} summarizes the established results and relations on which the analysis rests.  The
rest of the paper connects the two lines of work above through a property of single-scale unitarized amplitudes that
does not depend on the parameters, and uses it to ask which transverse structures are compatible with the data.

The analysis rests on a scale property of single-scale unitarized amplitudes (Sect.~\ref{sec:scale}).  For any
opacity with a fixed transverse shape and any unitarization map, the elastic fraction $r$,
$W=\sigma_{\rm tot}|t_{\rm dip}|/(8\pi)$ and $\Tbd=|t_{\rm bump}|/|t_{\rm dip}|$ depend on the central opacity alone,
whatever the energy dependence of the parameters, so that the amplitude is confined to one curve in $(r,W,\Tbd)$; for
the Gaussian Regge--eikonal this curve is given in closed form and $W\le3.75$.  Confronted with the measurements
(Sect.~\ref{sec:data}), the bound is violated at all eight energies from 23.5~GeV to 13~TeV at which the dip has been
located, by 7.6 to 71 standard deviations, and the energy trend of $W$ in the data is opposite to that of the model.
The energy-independent value of $\Tbd$ fixes the opacity, and since $r$ and $\Tbd$ are both monotonic in it, the rise
of $r$ from the ISR to the LHC excludes a single Gaussian scale independently of $W$.  The conflict is not removed by
the choice of the shape or of the unitarization (Sect.~\ref{sec:shape}): no stretched-exponential opacity with
$0.8\le\alpha\le6$ and no Gaussian $U$-matrix reaches the measured $(r,W,\Tbd)$, because long tails and reflective
centres raise $W$ but also $\Tbd$.  Two Gaussian scales reproduce $\sigma_{\rm tot}$, $r$, the forward slope and the dip
position at 13~TeV, but with the signature phase the position and the height of the bump cannot be matched together
(Sect.~\ref{sec:two}).  Finally, the parametrization commonly used to generate the phase violates $|S(b)|\le1$ beyond
an impact parameter $b_*$ given in closed form (Sect.~\ref{sec:tail}).  Table~\ref{tab:summary} lists the tests and their outcome.

The failure of a one-channel eikonal at the dip was noted qualitatively in
Ref.~\cite{PancheriGS,PancheriGS2}; here the
statement is generalized to arbitrary fixed shapes and unitarization maps and tested over the full energy range.  The
constancy of $\Tbd$ and of $\sigma_{\rm tot}|t_{\rm dip}|$ was established empirically and interpreted through geometric
scaling~\cite{BaldenegroPRS,Praszalowicz,Praszalowicz2,Praszalowicz3,Praszalowicz4,CsorgoNemes,CsorgoNemes2}; here these regularities are turned into bounds on the
transverse structure of unitarized Pomeron exchange.  L\'evy expansions were used to image the data~\cite{CsorgoLevy}
and eikonal and $U$-matrix schemes were compared in the forward region~\cite{Luna2026}; here both are confronted with
the dip--bump region through parameter-free combinations.  To our knowledge, the $\Tbd$ test, the exclusion of the
stretched-exponential family and of the Gaussian $U$-matrix, the bump tension of two-scale opacities and the closed form
of $b_*$ have not been given before.

\begin{table*}[t]
\caption{Summary of the parameter-free tests of this paper.  Deviations are in units of the experimental
uncertainty; $r$ is fixed to its measured value where indicated.}
\label{tab:summary}
\centering\footnotesize\setlength{\tabcolsep}{3pt}
\begin{tabular}{llccc}
\toprule
Amplitude class & Quantity & Model & Data & Deviation\\
\midrule
Gaussian eikonal, any parameters & $W$, 23.5~GeV--13~TeV & $\le3.75$ & 5.3--6.0 & 7.6--71$\sigma$\\
Gaussian eikonal, Eq.~\eqref{eq:born} & $W$, ISR$\to$LHC & rises, $2.9\to3.5$ & falls by $0.50\pm0.09$ & opposite trend\\
Gaussian eikonal, $r=0.281$ (13~TeV) & $\Tbd$ & 1.49 & $1.355\pm0.011$ & 12$\sigma$\\
L\'evy eikonal, $0.8\le\alpha\le6$, $r=0.281$ & $\Tbd$ & $\ge1.47$ & $1.355\pm0.011$ & $\ge11\sigma$\\
L\'evy eikonal, $0.8\le\alpha\le6$, $r=0.175$ (ISR) & $W$ & $\le4.4$ & $5.88$ & excluded\\
Gaussian $U$-matrix, $r=0.281$ & $W$, $\Tbd$ & 4.17, 1.378 & 5.32, 1.355 & 6.8$\sigma$, 2.1$\sigma$\\
Two-scale Gaussian with phase, 13~TeV & $R$ at measured $\Tbd$ & 1.24 & $1.77\pm0.01$ & excluded\\
Complex Regge slope & $|S(b)|\le1$ & violated for $b>b_*$ & -- & $b_*=5.8$--7.3~fm\\
\bottomrule
\end{tabular}
\end{table*}

\section{Established results and key relations}
\label{sec:background}

This section collects the results on which the analysis rests, in the normalization used throughout.

\subsection{Unitarity, impact parameter and bounds}

We normalize the amplitude so that~\cite{PDGsoft}
\begin{equation}
\sigma_{\rm tot}=\frac{{\rm Im}\,A(s,0)}{s},\qquad\frac{\dd\sigma_{\rm el}}{\dd t}=\frac{|A|^2}{16\pi s^2}.
\label{eq:bg-norm}
\end{equation}
The high-energy amplitude is the two-dimensional Fourier transform of the profile function $\Gamma(s,b)$,
\begin{equation}
A(s,t)=2is\int\dd^2b\,e^{i\q\cdot\bb}\,\Gamma(s,b),\qquad t=-\q^2,
\label{eq:bg-profile}
\end{equation}
where the impact parameter is related to the angular momentum by $b=2\ell/\sqrt s$~\cite{Collins}, so that
$S(s,b)=1-\Gamma(s,b)$ is the partial-wave $S$-matrix element.  Unitarity, $|S|\le1$, reads at each $b$
\begin{equation}
2\,{\rm Re}\,\Gamma=|\Gamma|^2+G_{\rm inel},\qquad 0\le G_{\rm inel}\le1,
\label{eq:bg-unit}
\end{equation}
and the integrated cross sections are
\begin{equation}
\begin{gathered}
\sigma_{\rm tot}=2\!\int\!\dd^2b\,{\rm Re}\,\Gamma,\qquad\sigma_{\rm el}=\int\!\dd^2b\,|\Gamma|^2,\\
\sigma_{\rm inel}=\int\!\dd^2b\,G_{\rm inel}.
\end{gathered}
\label{eq:bg-sigmas}
\end{equation}
For a purely absorptive profile, $0\le\Gamma\le1$; the black disk, $\Gamma=\theta(R-b)$, gives
\begin{equation}
\begin{gathered}
\sigma_{\rm tot}=2\pi R^2,\qquad\sigma_{\rm el}=\sigma_{\rm inel}=\pi R^2,\\
\frac{\dd\sigma_{\rm el}}{\dd t}=\frac{\pi R^4}{4}\left[\frac{2J_1(qR)}{qR}\right]^2,
\end{gathered}
\label{eq:bg-bd}
\end{equation}
with zeros at $qR=j_{1,n}$, so that $\sigma_{\rm el}/\sigma_{\rm tot}=1/2$ and $\sigma_{\rm tot}|t_{\rm dip}|=2\pi j_{1,1}^2$.
Values $1<\Gamma\le2$ are allowed by unitarity with $S<0$, the reflective regime~\cite{TroshinTyurin,TroshinTyurin2}.  Analyticity
and unitarity with a mass gap give the Froissart--Martin bound~\cite{Froissart,Martin}
\begin{equation}
\sigma_{\rm tot}(s)\le\frac{\pi}{m_\pi^2}\ln^2\frac{s}{s_0},
\label{eq:bg-froissart}
\end{equation}
whose coefficient, about 60~mb, is far above the coefficient 0.24--0.27~mb of $\ln^2s$ fits to the
data~\cite{PDGsoft}, and the Pomeranchuk theorem states that $\sigma_{\rm tot}^{pp}-\sigma_{\rm tot}^{p\bar p}\to0$ if
the ratio of real to imaginary parts grows more slowly than $\ln s$~\cite{Poghosyan}.

\subsection{Regge poles, the Pomeron and the real part}

The exchange of a Regge pole with trajectory $\alpha(t)$ and signature $\sigma=\pm1$ gives
\begin{equation}
A(s,t)=\eta_\sigma(t)\,\beta(t)\left(\frac{s}{s_0}\right)^{\alpha(t)},\quad
\eta_\sigma=-\frac{\sigma+e^{-i\pi\alpha}}{\sin\pi\alpha},
\label{eq:bg-regge}
\end{equation}
so that $\sigma_{\rm tot}\propto s^{\alpha(0)-1}$ and, for $\alpha(t)=\alpha(0)+\alpha't$ and an exponential residue,
$\dd\sigma_{\rm el}/\dd t\propto s^{2\alpha(0)-2}e^{B(s)t}$ with the shrinking slope
\begin{equation}
B(s)=B_0+2\alpha'\ln(s/s_0)
\label{eq:bg-shrink}
\end{equation}
\cite{Collins,Poghosyan}.  Secondary trajectories ($f,\omega,\rho,a_2$) have $\alpha(0)\simeq0.4$--0.7 and
$\alpha'\simeq0.8$--$0.9\GeVm$~\cite{Poghosyan}.  The rise of the total cross section requires the Pomeron, with
$\alpha_P(0)=1+\Delta>1$; the Donnachie--Landshoff fit~\cite{DL}
\begin{equation}
\sigma_{\rm tot}=X\,s^{\epsilon}+Y\,s^{-\eta},\qquad\epsilon=0.0808,\quad\eta=0.4525,
\label{eq:bg-DL}
\end{equation}
with $\alpha'_P=0.25\GeVm$~\cite{PDGsoft}, describes $pp$ and $p\bar p$ data up to the Tevatron with a common Pomeron
term and different secondary terms~\cite{DL,Poghosyan}.  Because the phase of an analytic amplitude is fixed by its
energy dependence, the forward ratio obeys the derivative dispersion relation
\begin{equation}
\rho(s)\simeq\frac{\pi}{2}\,\frac{\dd\ln\sigma_{\rm tot}}{\dd\ln s},\qquad
\rho=\tan\frac{\pi\epsilon}{2}\ \ \text{for}\ \ \sigma_{\rm tot}\propto s^\epsilon,
\label{eq:bg-ddr}
\end{equation}
which is the basis of the argument of Refs.~\cite{DL2019,DL2022} that $\rho$ cannot be extracted from a single
energy.  A $C$-odd exchange enters with opposite signs in the two channels,
$A^{pp}=A_++A_-$ and $A^{p\bar p}=A_+-A_-$; the Odderon is its leading-intercept member~\cite{Poghosyan,RyskinOdd}.

\subsection{Multiple exchange, eikonal and diffraction}

Single Pomeron exchange with $\Delta>0$ violates Eq.~\eqref{eq:bg-froissart}, and multi-Pomeron exchanges, the Regge
cuts of Gribov's calculus~\cite{Gribov}, must be added.  In the eikonal, which follows from the resummation of
Ref.~\cite{LevySucher} and from Glauber's multiple-scattering theory~\cite{Glauber},
\begin{equation}
\begin{gathered}
\Gamma=1-e^{-\Omega/2},\\
\Omega(s,b)=\frac{1}{s}\int\frac{\dd^2q}{(2\pi)^2}\,e^{-i\q\cdot\bb}\,\frac{A_B(s,-\q^2)}{i},
\end{gathered}
\label{eq:bg-eik}
\end{equation}
which corresponds to the $n$-Pomeron vertices of the Reggeon calculus taken equal to one~\cite{Poghosyan}; the
total cross section then grows as $\ln^2s$.  In a $U$-matrix scheme the same Born input is unitarized as
$\Gamma=(\Omega/2)/(1+\Omega/4)$, which agrees with Eq.~\eqref{eq:bg-eik} to second order but allows
$\Gamma>1$~\cite{Luna2026,Vanthieghe}.  Diffractive excitation is included by the Good--Walker
decomposition $|p\rangle=\sum_ic_i|\phi_i\rangle$ into eigenstates of diffraction~\cite{GoodWalker}, for which
\begin{equation}
\frac{\dd\sigma_{\rm el}}{\dd^2b}=|\langle\Gamma\rangle|^2,\qquad
\frac{\dd\sigma_{\rm el+D}}{\dd^2b}=\langle|\Gamma|^2\rangle,
\label{eq:bg-gw}
\end{equation}
so that low-mass diffraction measures the dispersion of the eigenvalues, and $\langle|\Gamma|^2\rangle\le{\rm Re}\langle
\Gamma\rangle$ gives the Pumplin bound $\sigma_{\rm el}+\sigma_{\rm D}\le\sigma_{\rm tot}/2$~\cite{Pumplin}.  The
Abramovsky--Gribov--Kancheli rules relate the discontinuities of an $n$-Pomeron graph with $k$ cut Pomerons,
$\propto(-1)^{n-k}2^n\binom{n}{k}$, so that the contributions of absorptive corrections cancel in inclusive
spectra~\cite{AGK,Poghosyan}.

\subsection{The diffractive dip and geometric scaling}

In geometric models the opacity is related to the hadronic matter distribution; the model of Chou and Yang takes it
proportional to the overlap of the electromagnetic form factors of the two protons~\cite{ChouYang}, and the eikonal
then produces diffraction zeros.  Empirically, the dip--bump region is often described by two exponentials with a
relative phase~\cite{PhillipsBarger},
\begin{equation}
\frac{\dd\sigma_{\rm el}}{\dd t}=\Big|\sqrt{A}\,e^{Bt/2}+\sqrt{C}\,e^{Dt/2+i\phi}\Big|^2,
\label{eq:bg-PB}
\end{equation}
which shows that the depth of the dip is controlled by the relative phase, i.e.\ by the real part of the
amplitude.  Geometric scaling~\cite{DiasDeDeus,BurasDiasDeDeus} is the hypothesis that the profile depends on $b$ only through
$b/R(s)$; with $\rho$ neglected it gives
\begin{equation}
\frac{\dd\sigma_{\rm el}}{\dd t}=\sigma_{\rm tot}^2\,\Phi(\tau),\qquad\tau=\sigma_{\rm tot}|t|,
\label{eq:bg-GS}
\end{equation}
with an energy-independent $\Phi$, so that $\sigma_{\rm el}/\sigma_{\rm tot}$, $B/\sigma_{\rm tot}$ and
$\sigma_{\rm tot}|t_{\rm dip}|$ are constant.  Geometric scaling holds at the ISR~\cite{AmaldiSchubert,DiasDeDeus,BurasDiasDeDeus},
whereas at the LHC $\sigma_{\rm el}/\sigma_{\rm tot}$ rises.  The product $\sigma_{\rm tot}|t_{\rm dip}|$ is
approximately constant from the ISR to the LHC but differs from the black-disk value $2\pi j_{1,1}^2=35.9\GeVt\,$mb
\cite{Bautista,CsorgoNemes,CsorgoNemes2}; the combination $|t_{\rm dip}|\sqrt{\sigma_{\rm el}\sigma_{\rm tot}}$ has been proposed
as a better invariant~\cite{PancheriGS,PancheriGS2,KMY,KMY2}; and the ratio of the bump to the dip position,
\begin{equation}
\Tbd=\frac{|t_{\rm bump}|}{|t_{\rm dip}|}=1.355\pm0.011,
\label{eq:bg-Tbd}
\end{equation}
is energy independent from 23~GeV to 13~TeV, which restores geometric scaling in the dip--bump region with
$|t_{\rm dip,bump}|\propto1/R^2(s)$ and, through crossing, gives $\rho=(\pi/2)\,\dd\ln R^2/\dd\ln s$
\cite{BaldenegroPRS,Praszalowicz,Praszalowicz2,Praszalowicz3,Praszalowicz4}.  In impact-parameter space the 13~TeV profile reconstructed with the method of
Ref.~\cite{AmaldiSchubert} is black at the centre, ${\rm Re}\,\Gamma(0)\simeq1.02$~\cite{Alkin}; L\'evy
imaging finds a small central depression of the inelastic profile~\cite{CsorgoLevy}, whose size depends on the $t$
dependence of the phase~\cite{Hollow}.

\subsection{Measurements}

Table~\ref{tab:meas} collects the LHC measurements used here and in the models discussed above.  The TOTEM and ATLAS
total cross sections differ by about 5\%, a two-standard-deviation difference that has been analysed in
Refs.~\cite{ATLAS13,PetrovATLAS,PetrovATLAS2}.  The forward ratio at 13~TeV, $\rho\simeq0.09$--0.10~\cite{TOTEMrho,ATLAS13}, lies
below the value $\simeq0.14$ obtained from the energy dependence~\cite{DL2019,DL2022}, which is the origin of the
Odderon discussion~\cite{MartynovNicolescu,D0TOTEM,LunaRyskinKhoze,LunaRyskinKhoze2,CsorgoUniverse,RyskinOdd}.  The differential cross
section is non-exponential already at small $|t|$~\cite{TOTEM8}, the forward slope grows faster than linearly in $\ln s$ above a few TeV, as discussed in
Refs.~\cite{TOTEMdsdt,KMRslope}, and the dip moves from about $1.4\GeVt$ at the ISR to $0.47\GeVt$ at 13~TeV while the
bump-to-dip cross-section ratio approaches about 1.8~\cite{TOTEMdsdt,BaldenegroPRS}.

\begin{table*}[t]
\caption{LHC measurements of the total cross section, the forward ratio, the dip position and the bump-to-dip
cross-section ratio $R$.  TOTEM values at 7 and 8~TeV are luminosity-independent; dashes mark quantities not
measured or not used here.}
\label{tab:meas}
\centering\footnotesize\setlength{\tabcolsep}{8pt}
\begin{tabular}{lccccc}
\toprule
$\sqrt s$ & $\sigma_{\rm tot}^{\rm TOTEM}$ [mb] & $\sigma_{\rm tot}^{\rm ATLAS}$ [mb] & $\rho$ & $|t_{\rm dip}|$ [GeV$^2$] & $R$\\
\midrule
2.76 TeV & $84.7\pm3.3$~\cite{TOTEMsig} & -- & -- & $0.61\pm0.03$~\cite{TOTEM276} & $1.7\pm0.2$~\cite{TOTEM276}\\
7 TeV & $98.0\pm2.5$~\cite{TOTEM7} & $95.35\pm1.36$~\cite{ATLAS7} & -- & $0.542\pm0.013$~\cite{BaldenegroPRS} & --\\
8 TeV & $101.7\pm2.9$~\cite{TOTEM8PRL} & $96.07\pm0.92$~\cite{ATLAS8} & -- & $0.521\pm0.007$~\cite{TOTEM8dip} & --\\
13 TeV & $110.6\pm3.4$~\cite{TOTEMsig} & $104.68\pm1.08$~\cite{ATLAS13} & 0.09--0.10~\cite{TOTEMrho} & $0.47\pm0.011$~\cite{TOTEMdsdt} & $1.77\pm0.01$~\cite{TOTEMdsdt}\\
\bottomrule
\end{tabular}
\end{table*}

\section{Single-scale unitarized amplitudes}
\label{sec:model}

With the normalization of Sect.~\ref{sec:background}, a
Pomeron pole with trajectory $\alpha_P(t)=1+\Delta+\alpha'_Pt$, an exponential residue and the signature factor
replaced by $i$ gives the Born amplitude~\cite{Collins,Poghosyan}
\begin{equation}
A_B(s,t)=i\,s\,H(s)\,e^{\lambda(s)t},\quad H\propto s^{\Delta},\quad\lambda=\lambda_0+\alpha'_P\ln s .
\label{eq:born}
\end{equation}
In the representation~\eqref{eq:bg-profile}, which we label
\begin{equation}
A(s,t)=2is\int\dd^2b\,e^{i\q\cdot\bb}\,\Gamma(s,b),
\label{eq:eik}
\end{equation}
for later reference, we use the eikonal profile $\Gamma=1-e^{-\Omega/2}$ or the $U$-matrix profile $\Gamma=(\Omega/2)/(1+\Omega/4)$, which agree
to second order in the opacity~\cite{Luna2026}.  For Eq.~\eqref{eq:born} the opacity is Gaussian,
$\Omega=2a\,e^{-b^2/(4\lambda)}$ with $a=H/(8\pi\lambda)$.  For the eikonal, expanding in the opacity and
integrating term by term gives, with $Q=\lambda|t|$,
\begin{equation}
A=8\pi is\lambda\,F(Q,a),\quad F(Q,a)=\sum_{n=1}^\infty\frac{(-1)^{n+1}a^n}{n!\,n}\,e^{-Q/n},
\label{eq:F}
\end{equation}
and, through the entire exponential integral $\Ein(a)=\int_0^a(1-e^{-u})u^{-1}\dd u$,
\begin{equation}
\begin{gathered}
\sigma_{\rm tot}=8\pi\lambda\Ein(a),\qquad\sigma_{\rm inel}=4\pi\lambda\Ein(2a),\\
r(a)=1-\frac{\Ein(2a)}{2\Ein(a)} .
\end{gathered}
\label{eq:sigmas}
\end{equation}
These closed forms follow from standard steps and are stated without a claim of priority.  The $n$-th term of Eq.~\eqref{eq:F} is the exchange of $n$ Pomerons.
At LHC opacities, $a\simeq2$--2.5, the terms are largest at $n=2$--3, and fixed-order truncations are not
approximations to the resummed amplitude: for $a=2.53$ the two-Pomeron truncation gives a $\sigma_{\rm tot}$ lower
by 39\% and its zero at $0.124\GeVt$ instead of $0.315\GeVt$, the three-Pomeron truncation has no zero, and the zero
converges only beyond eight Pomerons (Eq.~\eqref{eq:F} truncated at fixed $n$).

\section{Scale invariance and the bounds}
\label{sec:scale}

Let the opacity have a fixed shape, $\Omega(b)=2a\,\phi\big(b^2/(4\lambda)\big)$ with $\phi(0)=1$, and let
$\Gamma=g(\Omega)$ for a fixed unitarization map $g$.  With $z=b^2/(4\lambda)$, Eq.~\eqref{eq:eik} becomes
\begin{equation}
\frac{A}{8\pi is\lambda}=\int_0^\infty\dd z\,J_0\big(2\sqrt{Qz}\big)\,g\big(2a\phi(z)\big),
\label{eq:scale}
\end{equation}
so that $\sigma_{\rm tot}/\lambda$, $\sigma_{\rm el}/\lambda$ and the zeros and maxima of $A$ in $Q$ are functions of
$a$ alone.  Every dimensionless combination of these quantities, in particular
\begin{equation}
\begin{gathered}
W(a)=\frac{\sigma_{\rm tot}|t_0|}{8\pi},\qquad \Tbd(a)=\frac{Q_1(a)}{Q_0(a)},\qquad r(a),
\end{gathered}
\label{eq:W}
\end{equation}
where $Q_0$ is the first zero of $A$ and $Q_1$ the following maximum of $|A|^2$, depends on $a$ only, whatever the
energy dependence of $a$ and $\lambda$.  A single-scale amplitude is therefore confined to one curve in the space of
$(r,W,\Tbd)$, fixed by the pair $(\phi,g)$.  For the black disk, the limit $a\to\infty$ of the eikonal, the amplitude
is proportional to $J_1(qR)/q$ and gives $r=1/2$, $W=j_{1,1}^2/4=3.67$ and $\Tbd=(j_{2,1}/j_{1,1})^2=1.796$, with
$j_{1,1}=3.8317$ the first zero of $J_1$ and $j_{2,1}=5.1356$ the first maximum of $|J_1(x)/x|$ beyond
it~\cite{DLMF}; $W_{\rm BD}$ corresponds to the value $35.9\GeVt\,$mb of Ref.~\cite{Bautista}.

\begin{figure}[t]
\centering\includegraphics[width=\columnwidth]{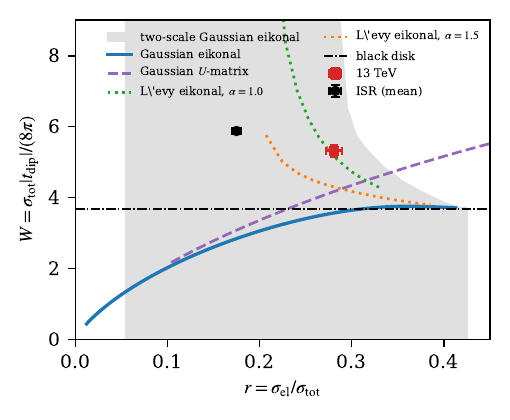}
\caption{The curves $(r,W)$ of single-scale amplitudes: Gaussian eikonal, Eqs.~\eqref{eq:sigmas}
and~\eqref{eq:W}; Gaussian $U$-matrix; L\'evy eikonal with $\alpha=1$ and 1.5 (Sect.~\ref{sec:shape}); the region
reached by real two-scale Gaussian eikonals (upper envelope of the scan of Sect.~\ref{sec:two}, truncated at
$W=14$); the black-disk value; TOTEM at 13~TeV~\cite{TOTEMsig,BaldenegroPRS}; and the weighted ISR average of
Table~\ref{tab:multiW} at $r\simeq0.175$~\cite{AmaldiSchubert}.}
\label{fig:rW}
\end{figure}

For the Gaussian eikonal, $W(a)=\Ein(a)Q_0(a)$ rises from 2.81 at $a=1$ to a maximum 3.75 at $a=6.8$, where
$r=0.361$, and then approaches the black-disk value (Fig.~\ref{fig:rW}); $\Tbd(a)$ rises monotonically, through
1.24, 1.355, 1.49 and 1.68 at $a=0.3$, 1.0, 2.53 and 12, towards 1.796 (Fig.~\ref{fig:rT}).  Hence, for any
parameters,
\begin{equation}
W\le3.75 ,
\label{eq:bound}
\end{equation}
and $\Tbd$ determines $a$ and therefore $r$ and $W$.  Identifying the measured dip with the zero is not affected by
the real part: restoring the signature phase with the trajectory of Ref.~\cite{DL}, $\Delta=0.0808$ and
$\alpha'_P=0.25\GeVm$, through the complex slope of Ref.~\cite{Poghosyan}, fills the zero of the 13~TeV calibration
($a=2.45$, $\lambda=7.52\GeVm$) with $\rho=0.098$ and moves the minimum from 0.315 to $0.317\GeVt$.

\section{Confrontation with data from the ISR to the LHC}
\label{sec:data}

\begin{figure}[t]
\centering\includegraphics[width=\columnwidth]{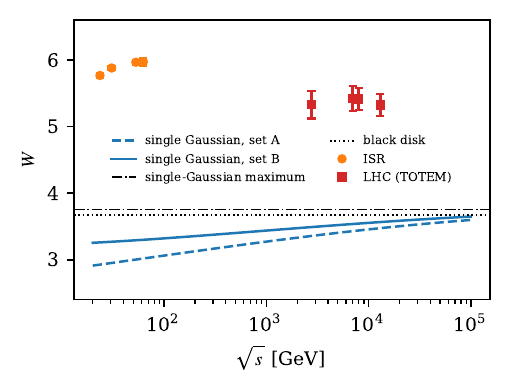}
\caption{$W(\sqrt s)$ from Table~\ref{tab:multiW}, the Gaussian-eikonal maximum 3.75, the black-disk value 3.67, and
the Gaussian eikonal evolved with the trajectory of Ref.~\cite{DL} for sets A and B.}
\label{fig:Ws}
\end{figure}

Table~\ref{tab:multiW} and Fig.~\ref{fig:Ws} give $W$ at the eight energies at which the dip has been located, from
the ISR values of $\sigma_{\rm tot}$ compiled in Ref.~\cite{NemesCsorgo2012}, the TOTEM values at 2.76, 7, 8 and
13~TeV~\cite{TOTEMsig,TOTEM7,TOTEM8PRL} and the dip positions of Refs.~\cite{BaldenegroPRS,TOTEM8dip}, with the
uncertainties of $\sigma_{\rm tot}$ and $|t_{\rm dip}|$ added in quadrature.  All values lie between 5.3 and 6.0,
7.6 to 71 standard deviations above Eq.~\eqref{eq:bound}; the ISR significances do not include a systematic
uncertainty of the dip definition and are upper limits.  Extracting $\sigma_{\rm tot}$ at 13~TeV with
$\rho=0.14$~\cite{DL2022} instead of 0.10 lowers it by 0.94\% and gives $W=5.26\pm0.21$, still 7.4 standard
deviations above the bound; the ATLAS normalization~\cite{ATLAS13}, lower by about 5\%, lowers $W$ by the same
fraction.

The energy trend is opposite to that of the model.  Evolved with Eq.~\eqref{eq:born} and the trajectory of
Ref.~\cite{DL}, from a normalization at 7~TeV (set A, $\lambda=10\GeVm$, $a=2$) or from the 13~TeV calibration to
$\sigma_{\rm tot}$ and $r$ (set B, $a=2.53$, $\lambda=7.38\GeVm$), the opacity grows from $a=1.1$ and
1.6 at 23.5~GeV to 2.1 and 2.5 at 13~TeV, so that the model $W$ rises from 2.9 and 3.3 to 3.5 and 3.6.  The data
fall from a weighted ISR average of $5.876\pm0.018$ to an LHC average of $5.375\pm0.091$, a decrease of
$0.50\pm0.09$; within the LHC range they are consistent with a constant, $\dd W/\dd\ln s=-0.0005\pm0.086$, in
agreement with the nearly equal powers of $\sigma_{\rm tot}$ and $|t_{\rm dip}|$ found in
Refs.~\cite{BaldenegroPRS,Praszalowicz,Praszalowicz2,Praszalowicz3,Praszalowicz4}.

\begin{figure}[t]
\centering\includegraphics[width=\columnwidth]{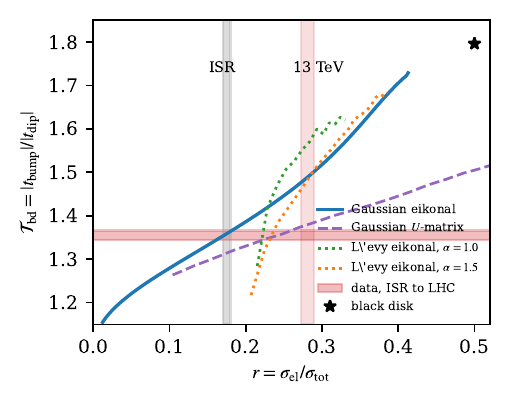}
\caption{The curves $(r,\Tbd)$ of the Gaussian eikonal, the Gaussian $U$-matrix and the L\'evy eikonal with $\alpha=1$
and 1.5, the measured $\Tbd=1.355\pm0.011$~\cite{BaldenegroPRS}, the elastic fractions at the ISR~\cite{AmaldiSchubert}
and at 13~TeV~\cite{TOTEMsig}, and the black-disk point.}
\label{fig:rT}
\end{figure}

The ratio $\Tbd$ gives an independent test.  The measured value fixes $a=1.0$, for which $r=0.172$ and $W=2.81$.  The
elastic fraction of this solution is close to that of the ISR data, $r\simeq0.17$--0.18~\cite{AmaldiSchubert}, so
that at the ISR a single Gaussian reproduces $r$ and $\Tbd$ together but not $W$ (Fig.~\ref{fig:rT}).  At 13~TeV,
$r=0.281\pm0.009$~\cite{TOTEMsig} requires $\Tbd=1.49$, twelve standard deviations above the measured ratio.  Since $r$
and $\Tbd$ are both monotonic in $a$, an energy-independent $\Tbd$ with a rising $r$ cannot be reproduced by a
single-scale Gaussian eikonal for any energy dependence of its parameters.  In the language of geometric scaling,
a single-scale amplitude scales only at fixed $a$; the data keep the dip--bump pattern fixed in units of
$|t_{\rm dip}|$ while the elastic fraction changes, so that the shape of the profile, and not only its size, must
change with energy.

\begin{table}[t]
\caption{$W=\sigma_{\rm tot}|t_{\rm dip}|/(8\pi)$ from the measured $\sigma_{\rm tot}$ and dip position, its deviation
from the maximum 3.75 of Eq.~\eqref{eq:bound} in units of its uncertainty, and $W(a)$ of the Gaussian eikonal
evolved with the trajectory of Ref.~\cite{DL} for sets A and B.}
\label{tab:multiW}
\centering\scriptsize\setlength{\tabcolsep}{2.2pt}
\begin{tabular}{lcccccc}
\toprule
$\sqrt s$ & $\sigma_{\rm tot}$ [mb] & $|t_{\rm dip}|$ [GeV$^2$] & $W$ & dev. & $W_A$ & $W_B$\\
\midrule
23.5 GeV & $38.94\pm0.17$ & $1.450\pm0.005$ & $5.77\pm0.03$ & 63 & 2.93 & 3.26\\
30.6 GeV & $40.14\pm0.17$ & $1.434\pm0.004$ & $5.88\pm0.03$ & 71 & 2.95 & 3.27\\
52.8 GeV & $42.67\pm0.19$ & $1.369\pm0.006$ & $5.97\pm0.04$ & 59 & 3.00 & 3.29\\
62.5 GeV & $43.32\pm0.23$ & $1.350\pm0.011$ & $5.98\pm0.06$ & 38 & 3.02 & 3.30\\
2.76 TeV & $84.7\pm3.3$ & $0.616\pm0.002$ & $5.33\pm0.21$ & 7.6 & 3.36 & 3.49\\
7 TeV & $98.0\pm2.5$ & $0.542\pm0.013$ & $5.43\pm0.19$ & 8.8 & 3.43 & 3.54\\
8 TeV & $101.7\pm2.9$ & $0.521\pm0.007$ & $5.41\pm0.17$ & 9.7 & 3.44 & 3.54\\
13 TeV & $110.6\pm3.4$ & $0.471\pm0.003$ & $5.32\pm0.17$ & 9.4 & 3.47 & 3.57\\
\bottomrule
\end{tabular}
\end{table}

\section{Other shapes and other unitarizations}
\label{sec:shape}

\begin{figure}[t]
\centering\includegraphics[width=\columnwidth]{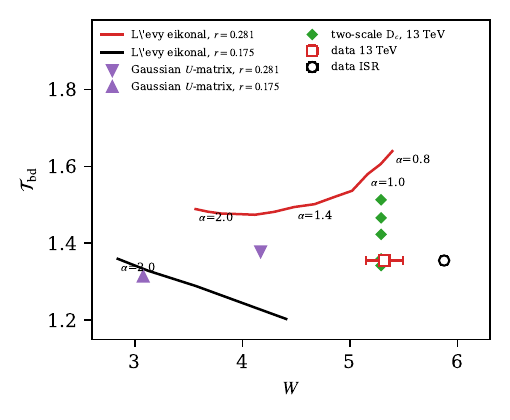}
\caption{$(W,\Tbd)$ at fixed elastic fraction: the L\'evy eikonal for $0.8\le\alpha\le2$ at $r=0.281$ (13~TeV) and for
$1.7\le\alpha\le2$ at $r=0.175$ (ISR; smaller $\alpha$ give no zero), the Gaussian $U$-matrix at both values of $r$,
the two-scale family D$_c$ of Sect.~\ref{sec:two} at 13~TeV, and the data.}
\label{fig:WT}
\end{figure}

The bound~\eqref{eq:bound} is specific to the Gaussian, but Eq.~\eqref{eq:scale} applies to every single-scale
amplitude.  To test whether a different shape can reach the data, we consider the stretched-exponential (L\'evy)
opacities $\phi_\alpha(z)=\exp(-z^{\alpha/2})$, i.e.\ $\Omega=2a\exp[-(b/b_0)^\alpha]$, which contain the Gaussian at
$\alpha=2$, longer tails for $\alpha<2$ and flatter, more disk-like profiles for $\alpha>2$, and are used for the
model-independent imaging of the 13~TeV data~\cite{CsorgoLevy}.  At fixed $r$ each $\alpha$ determines $a$, $W$ and
$\Tbd$.  At the 13~TeV value $r=0.281$, $W$ decreases monotonically from 5.39 at $\alpha=0.8$ through 5.16 at
$\alpha=1$ and 3.57 at $\alpha=2$ to 2.49 at $\alpha=6$, so that the measured $W$ requires $\alpha\simeq0.8$--1.0;
but $\Tbd$ is then 1.58--1.64, and its minimum over $0.8\le\alpha\le6$ is 1.47, reached at $\alpha\simeq1.6$, eleven
standard deviations above the measured 1.355 (Fig.~\ref{fig:WT}).  At the ISR value $r=0.175$, shapes with
$\alpha<1.7$ produce no zero at all, and for $\alpha\ge1.7$ they give $W\le4.4$, against $5.88$.  Long tails thus
raise $W$ but also $\Tbd$, while the data require a large $W$ and a small $\Tbd$ at the same time; no member of this
family of single-scale shapes reaches the measured $(r,W,\Tbd)$ at either energy.

The unitarization map matters more.  With the same Gaussian opacity, the $U$-matrix profile
$\Gamma=(\Omega/2)/(1+\Omega/4)$ allows $\Gamma>1$, the reflective regime~\cite{TroshinTyurin,TroshinTyurin2}.  At $r=0.281$ it gives
$a=2.03$, a central profile $\Gamma(0)=1.007$, in line with the black centre of the data~\cite{Alkin},
$\Tbd=1.378$, within $2.1\sigma$ of the measurement, and $W=4.17$, still $6.8\sigma$ below the measured value; at
$r=0.175$ it gives $W=3.08$ and $\Tbd=1.31$.  The comparison of the two maps in Figs.~\ref{fig:rW}--\ref{fig:WT}
shows that the dip--bump structure is sensitive to the unitarization scheme already for a single scale, a property
not visible in fits restricted to $|t|<0.1\GeVt$~\cite{Luna2026}, but neither map reaches $W$ with one scale.

\section{Two transverse scales}
\label{sec:two}

For the real opacity $\Omega=2a_1e^{-b^2/(4\lambda_1)}+2a_2e^{-b^2/(4\lambda_2)}$ the expansion of Eq.~\eqref{eq:F}
becomes a double series,
\begin{equation}
\begin{gathered}
\frac{A}{8\pi is}=\sum_{n+m\ge1}\frac{(-1)^{n+m+1}a_1^na_2^m}{n!\,m!}\,\lambda_{nm}e^{\lambda_{nm}t},\\
\lambda_{nm}^{-1}=\frac{n}{\lambda_1}+\frac{m}{\lambda_2},
\end{gathered}
\label{eq:dgseries}
\end{equation}
which reduces to Eq.~\eqref{eq:F} for $a_2=0$ or $\lambda_1=\lambda_2$, and $r$, $W$, $\Tbd$ depend on
$(a_1,a_2,k=\lambda_2/\lambda_1)$.  A scan of 84\,640 points with $0.03\le a_j\le25$ and $0.005\le k\le1$, checked
against the single-Gaussian maximum at $k=1$, reaches $W=12.6$ at $r=0.281\pm0.009$ (Fig.~\ref{fig:rW}).  The TOTEM
point lies inside the family, and all members consistent with it have a dense inner component, $k\le0.29$, and a
nearly black centre, $\Gamma(0)\ge0.978$.

\begin{figure}[t]
\centering\includegraphics[width=\columnwidth]{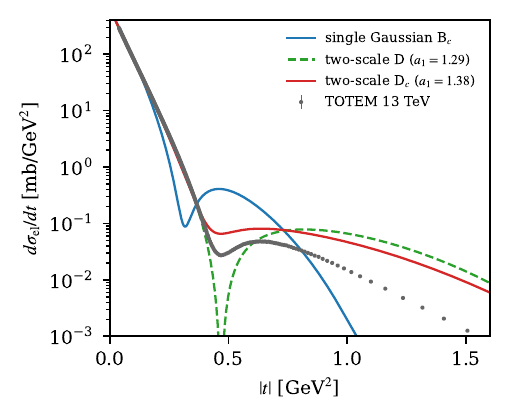}
\caption{Elastic differential cross section at 13~TeV~\cite{TOTEMdsdt} compared with the complex single Gaussian
calibrated to $\sigma_{\rm tot}$ and $r$ (B$_c$), the real two-scale member of set D with $a_1=1.29$, and the complex
two-scale member of set D$_c$ with $a_1=1.38$, which reproduces $\Tbd$.}
\label{fig:dsdt}
\end{figure}

Imposing $\sigma_{\rm tot}=110.6$~mb, $r=0.281$ and a zero at $|t_0|=0.468\GeVt$ at 13~TeV leaves a one-parameter
family, set D (Table~\ref{tab:D}).  Its member with $a_1=1.29$, $a_2=3.51$, $\lambda_1=9.70$ and
$\lambda_2=2.56\GeVm$ has a log-linear slope of $20.37\GeVm$ in $0.04\le|t|\le0.2\GeVt$, against $20.39\GeVm$ for
the data~\cite{TOTEMdsdt}, and describes 71\% of the bins below $1\GeVt$ within 10\%: a second scale reconciles
$\sigma_{\rm tot}$, $r$, the forward slope and the dip position.  It does not reconcile the bump, which lies at
$0.82\GeVt$ instead of $0.632\GeVt$ ($\Tbd=1.75$), nor the large-$|t|$ tail.  With the signature phase in both
components and the same three conditions (set D$_c$), the zero is filled with $\rho=0.102$--0.103, and along the
family $\Tbd$ falls from 1.51 at $a_1=1.0$ to 1.34 at $a_1=1.4$.  The measured 1.355 is reached at $a_1\simeq1.38$,
where the bump-to-dip cross-section ratio is $R=1.24$ and the model exceeds the data at the dip by a factor of 2.4;
the measured $R=1.77$~\cite{TOTEMdsdt} requires $a_1\simeq1.06$, where $\Tbd\simeq1.50$ (Figs.~\ref{fig:WT}
and~\ref{fig:dsdt}).  A two-scale Gaussian opacity constrained by $\sigma_{\rm tot}$, $r$ and the dip position
therefore cannot reproduce both the position and the height of the bump.  Because $R$ depends on the forward phase
roughly as $(1+\rho^2)/\rho^2$~\cite{Praszalowicz,Praszalowicz2,Praszalowicz3,Praszalowicz4}, this tension refers to the trajectory of Ref.~\cite{DL} and should
be re-examined with that of the multi-energy fit of Ref.~\cite{DL2022}, for which the single Gaussian gives
$\rho=0.131$.

\begin{table*}[t]
\caption{Two-scale opacities constrained to $\sigma_{\rm tot}=110.6$~mb, $r=0.281$ and a first minimum at
$0.468\GeVt$ at 13~TeV: real (set D) and with the signature phase (set D$_c$), for members of each one-parameter
family labelled by $a_1$.  Widths in GeV$^{-2}$; the slope is the equal-weight log-linear slope in
$0.04\le|t|\le0.2\GeVt$ (data: 20.39); ratios refer to bin-averaged model values over the data of
Ref.~\cite{TOTEMdsdt}.}
\label{tab:D}
\centering\footnotesize\setlength{\tabcolsep}{6pt}
\begin{tabular}{lccccccc}
\toprule
 & \multicolumn{3}{c}{D (real)} & \multicolumn{4}{c}{D$_c$ (complex)}\\
\cmidrule(lr){2-4}\cmidrule(lr){5-8}
$a_1$ & 1.0 & 1.29 & 1.4 & 1.0 & 1.29 & 1.38 & 1.4\\
\midrule
$a_2$ & 3.688 & 3.514 & 3.517 & 3.071 & 2.779 & 2.709 & 2.695\\
$\lambda_1,\lambda_2$ & 10.71, 2.95 & 9.70, 2.56 & 9.40, 2.38 & 10.71, 3.21 & 9.71, 2.83 & 9.47, 2.69 & 9.42, 2.66\\
$\rho$ & 0 & 0 & 0 & 0.103 & 0.102 & 0.102 & 0.102\\
slope, 0.04--0.2 & 19.80 & 20.37 & 20.55 & 20.15 & 20.72 & -- & 20.91\\
bump [GeV$^2$] & 0.790 & 0.818 & 0.838 & 0.708 & 0.666 & 0.636 & 0.628\\
$\Tbd$ & 1.69 & 1.75 & 1.79 & 1.51 & 1.42 & 1.36 & 1.34\\
bump/dip ratio $R$ & -- & -- & -- & 1.89 & 1.37 & 1.24 & 1.21\\
model/data at dip & -- & -- & -- & 2.81 & 2.50 & 2.41 & 2.39\\
bins within 10\%, $|t|<1$ & 38\% & 71\% & 63\% & 47\% & 69\% & -- & 49\%\\
\bottomrule
\end{tabular}
\end{table*}

\section{Pointwise unitarity of the complex Regge slope}
\label{sec:tail}

The phase of the Born amplitude is commonly generated by the complex slope $\lambda_c=\lambda-i\pi\alpha'_P/2$
together with $H_c=H(1-i\tan\frac{\pi\Delta}{2})$~\cite{Poghosyan}.  Since
$1/\lambda_c=(\lambda+i\pi\alpha'_P/2)/|\lambda_c|^2$, the phase of $\Omega(b)$ decreases as
$-\pi\alpha'_Pb^2/(8|\lambda_c|^2)$, and ${\rm Re}\,\Omega$, hence $1-|S(b)|^2$ in the eikonal, becomes negative for
\begin{equation}
b^2>b_*^2=\frac{8|\lambda_c|^2}{\pi\alpha'_P}\left[\frac{\pi(1-\Delta)}{2}
+\arctan\frac{\pi\alpha'_P}{2\lambda}\right],
\label{eq:tailonset}
\end{equation}
where $|\Omega|$ is of order $|\nu_c|\exp\{-(2\lambda/\pi\alpha'_P)[\pi(1-\Delta)/2+\arctan(\pi\alpha'_P/2\lambda)]\}$.
With 40-digit arithmetic, Eq.~\eqref{eq:tailonset} gives $b_*=5.80$~fm and $\min_b{\rm Re}\,\Omega=-3.5\times10^{-14}$
for the calibrated complex single Gaussian, and $b_*=7.28$~fm with $\min_b{\rm Re}\,\Omega=-1.2\times10^{-17}$ for the
D$_c$ member with $a_1=1.38$.  The violation does not affect any observable quoted here, but the parametrization
satisfies $|S|\le1$ only to this accuracy; the integrated identity $\sigma_{\rm tot}=\sigma_{\rm el}+\sigma_{\rm inel}$
holds for any complex $\Omega$ and does not test it.  A phase carried by the coupling alone, with real $\lambda$,
keeps ${\rm Re}\,\Omega\ge0$ everywhere, since $|\arg H_c|=\pi\Delta/2<\pi/2$, at the price of a $t$-independent
phase.

\section{Discussion}
\label{sec:disc}

The results separate what a profile must do to describe the dip region from what the parametrization happens to
allow.  A single transverse scale, whatever its shape within the stretched-exponential family and whichever of the
two unitarization maps is used, cannot give the measured $W$ and $\Tbd$ together: long tails or reflective centres
raise $W$ but also $\Tbd$, while a small $\Tbd$ requires a grey profile with a small $W$.  The data therefore require
at least two transverse scales in the dip--bump region, a dense core and a broader periphery, which is the structure
of the reconstructed profiles~\cite{Alkin,CsorgoLevy} and of the successful models, where it arises from
multichannel eigenstates, non-Gaussian form factors or composite proton
structure~\cite{KMR,KMRslope,HEGS,HEGS2,CsorgoNemes,CsorgoNemes2}.  Two Gaussian scales are not sufficient for the bump, which suggests
that the periphery must fall more slowly than a Gaussian, as for the power-like form factors of
Refs.~\cite{KMR,KMRslope}.

The energy dependence points in the same direction.  In a single-scale model the opacity grows with energy and $W$
rises towards its maximum, whereas the data fall from the ISR to the LHC and then stay constant.  Since $\Tbd$ is
constant while $r$ rises, the relative weight of core and periphery must change with energy in a way that keeps the
dip--bump pattern fixed in units of $|t_{\rm dip}|$; this is the regime in which geometric scaling holds in the
dip--bump region but not at small $|t|$~\cite{BaldenegroPRS,Praszalowicz,Praszalowicz2,Praszalowicz3,Praszalowicz4}.

The conclusions do not depend on the forward phase.  The real part fills the dip but moves it by less than one per
cent; replacing $\rho=0.10$ by the value 0.14 of Refs.~\cite{DL2019,DL2022} changes $W$ by one per cent; and the
ATLAS normalization changes $W$ by about five per cent.  The diagnostics are therefore insensitive to the Odderon
question at small $|t|$, which concerns the phase~\cite{DL2022,LunaRyskinKhoze,LunaRyskinKhoze2,RyskinOdd}, but they constrain the
transverse structure that any $C$-even or $C$-odd description of the dip must have.

The diagnostics can be applied in three ways.  The predictions of published multi-energy
models~\cite{KMR,HEGS,HEGS2,CsorgoUniverse} for $W(s)$, $\Tbd(s)$ and $R(s)$ test their transverse structure directly and do
not depend on their parametrization.  If $W$ remains constant at the LHC, $W=5.375\pm0.091$, a measurement of
$\sigma_{\rm tot}$ at a new energy fixes the dip position through $|t_{\rm dip}|=8\pi W/\sigma_{\rm tot}$ and the bump
through $\Tbd$, which the running at $\sqrt s=13.6$~TeV can check.  The same diagnostics applied to $p\bar p$ data, in
which the dip is replaced by a shoulder~\cite{D0TOTEM}, constrain the $C$-odd contribution to the profile.

\section{Conclusions}
\label{sec:concl}

In a unitarized amplitude with a fixed transverse shape, $r$, $W=\sigma_{\rm tot}|t_{\rm dip}|/(8\pi)$ and $\Tbd$
depend on one parameter, whatever the energy dependence of the parameters, and for the Gaussian Regge--eikonal these
functions are explicit with $W\le3.75$.  The data violate this bound at all eight energies from 23.5~GeV to 13~TeV, by
7.6 to 71 standard deviations, with an energy trend opposite to that of the model.  The energy independence of $\Tbd$,
together with the rise of $r$ from the ISR to the LHC, excludes a single Gaussian scale independently of $W$.
Stretched-exponential shapes with any exponent between 0.8 and 6 and the Gaussian $U$-matrix do not reach the measured
$(r,W,\Tbd)$ either, although the $U$-matrix reproduces $\Tbd$ and the black centre at 13~TeV, so that the data require
at least two transverse scales in the dip--bump region.  Two Gaussian scales reconcile $\sigma_{\rm tot}$, $r$, the
forward slope and the dip at 13~TeV, but not the position and the height of the bump together, which points to a
periphery falling more slowly than a Gaussian.  The complex Regge slope, in addition, violates $|S(b)|\le1$ beyond
$b_*$ of Eq.~\eqref{eq:tailonset}.  These parameter-free diagnostics do not depend on the parametrization or on the forward phase, and can be applied
directly to the predictions of any model of the dip region, in particular to the multi-energy models of
Refs.~\cite{KMR,HEGS,HEGS2,CsorgoUniverse}, and to future measurements at $\sqrt s=13.6$~TeV.

\appendix
\section{Numerical methods}
\label{app:num}

$W(a)$ and $\Tbd(a)$ of the Gaussian eikonal are computed from Eq.~\eqref{eq:F} summed to $n=200$, with $\Ein$ from
$\gamma_E+\ln a+E_1(a)$, the zero by Brent's method and the maximum by bounded maximization.  For the L\'evy and
$U$-matrix profiles and for two scales, Eq.~\eqref{eq:scale} is evaluated by composite 16-point Gauss--Legendre
quadrature on 300 panels in the variable $\sqrt z$, which reproduces Eq.~\eqref{eq:F} for the Gaussian eikonal to
$10^{-6}$ in $r$, $W$ and $\Tbd$; zeros are located on a grid of 3000--4000 points in $Q$ with linear interpolation.
The three conditions defining sets D and D$_c$ are solved by Powell's hybrid method, model values are averaged over
the bins of Ref.~\cite{TOTEMdsdt}, and Eq.~\eqref{eq:tailonset} and the minima of ${\rm Re}\,\Omega$ are evaluated with
40-digit arithmetic.

\end{document}